\documentclass[aps,prl,groupedaddress,superscriptaddress,twocolumn,notitlepage,nofootinbib
]{revtex4-2}
\pdfoutput=1

\usepackage{dcolumn}
\usepackage{physics,amsmath,amsthm,amsfonts,graphicx,times,xfrac,mathtools,mathrsfs,amssymb,bm,bbm,verbatim,appendix,mathrsfs,scalerel,lipsum}
\usepackage[normalem]{ulem}
\usepackage{xcolor,hyperref}
\usepackage[caption=false,labelfont={normalsize}]{subfig}
\theoremstyle{plain}
\newtheorem{definition}{Definition}
\newtheorem{lemma}{Lemma}

\newtheorem{theorem}{Theorem}

\newcommand{\Def}{\coloneqq}

\newcommand{\Sbb}{\mathbb{S}}
\newcommand{\Pbb}{\mathbb{P}}

\newcommand{\Acal}{\mathcal{A}}

\newcommand{\Dcal}{\mathcal{D}}
\newcommand{\Ecal}{\mathcal{E}}
\newcommand{\Fcal}{\mathscr{F}}
\newcommand{\Gcal}{\mathscr{G}}
\newcommand{\Hcal}{\mathcal{H}}
\newcommand{\Ical}{\mathcal{I}}

\newcommand{\Pcal}{\mathcal{P}}
\newcommand{\Qcal}{\mathcal{Q}}

\newcommand{\Scal}{\mathcal{S}}

\newcommand{\Vcal}{\mathcal{V}}
\newcommand{\Xcal}{\mathcal{X}}

\DeclareMathOperator*{\argmax}{argmax}

\begin{document}

\title{Quantum Processes as Thermodynamic Resources: The Role of Non-Markovianity}

\author{Guilherme Zambon}
\email{guilhermezambon@usp.br}
\affiliation{
 Instituto de F{\'i}sica de S{\~a}o Carlos, Universidade de S{\~a}o Paulo, CP 369, 13560-970, S{\~a}o Carlos, SP, Brasil
 }
 \affiliation{
 School of Mathematical Sciences and Centre for the Mathematical and Theoretical Physics of Quantum Non-Equilibrium Systems, University of Nottingham, University Park, Nottingham, NG7 2RD, United Kingdom
 }
\author{Gerardo Adesso}
\email{gerardo.adesso@nottingham.ac.uk}
\affiliation{
 School of Mathematical Sciences and Centre for the Mathematical and Theoretical Physics of Quantum Non-Equilibrium Systems, University of Nottingham, University Park, Nottingham, NG7 2RD, United Kingdom
 }

\begin{abstract}
  Quantum thermodynamics studies how quantum systems and operations may be exploited as sources of work to perform useful thermodynamic tasks. In real-world conditions, the evolution of open quantum systems typically displays memory effects, resulting in a non-Markovian dynamics. The associated information backflow has been observed to provide advantage in certain thermodynamic tasks. However, a general operational connection between non-Markovianity and thermodynamics in the quantum regime has remained elusive. Here, we analyze the role of non-Markovianity in the central task of extracting work via thermal operations from general multitime quantum processes, as described by process tensors.  By defining a hierarchy of four classes of extraction protocols, expressed as quantum combs, we reveal three different physical mechanisms (work investment, multitime correlations, and system-environment correlations) through which non-Markovianity increases the work distillable from the process. The advantages arising from these mechanisms are linked precisely to a quantifier of the non-Markovianity of the process. These results show in very general terms how non-Markovianity of any given quantum process is a fundamental resource that unlocks an enhanced performance in thermodynamics.
\end{abstract}

\maketitle

\textit{Introduction.}--- Designing efficient strategies to extract work has been the cornerstone of thermodynamics since the dawn of the first industrial revolution. The advent of \textit{ quantum thermodynamics} \cite{QTBook} has opened new scenarios to identify fundamental mechanisms, spanning from coherence and correlations to reservoir engineering and even ignorance \cite{Perarnau2015extractable,Korzekwa2016extraction,Francica2017daemonic,Morris2019assisted,Yadin2021mixing}, which may be exploited to extract additional work from quantum systems.  The modern formulation of quantum thermodynamics as a resource theory \cite{janzing2000thermodynamic,brandao2013resource,Ng2018,lostaglio2019introductory} in fact provides an operational justification for the focus on work extraction: given that the theory is asymptotically reversible both at static \cite{horodecki2013fundamental,brandao2013resource,brandao2015second} and dynamic \cite{navascues2015nonthermal,faist2019thermodynamic,faist2021thermodynamic} levels, any state or channel transformation is fully characterized in the asymptotic limit by its work content. We are thus faced with the key question: \textit{ How can we maximize work extraction from general quantum processes?}

When investigating quantum properties of general thermodynamic tasks, Markovianity of the underlying dynamics is often assumed for ease of analytical calculations, but such a simplification comes with a price to pay. That is, in a Markovian dynamics any resource the system loses to the environment at some point in time cannot be later recovered and consumed for the task. Conversely, non-Markovian dynamics allows for such a resource backflow \cite{bylicka2016thermodynamic}, which has been considered as a potential source of advantage in a number of thermodynamic tasks \cite{strasberg2016nonequilibrium,thomas2018thermodynamics,pezzutto2019out,abiuso2019non,kamin2020non,bhattacharya2020thermodynamic,debiossac2020thermodynamics,abah2020implications,camati2020employing,shirai2021non,halpern2020fundamental,spaventa2022capacity,ptaszy2022non,cheong2023effects,picatoste2024dynamically,mohammadi2024quantum}.
While for a continuous-time evolution these phenomena can be explored using tools such as non-Markovian master equations \cite{vacchini2013non,das2021thermodynamics,colla2022open}, dynamical maps \cite{rivas2020strong}, or continuous thermomajorization \cite{lostaglio2018elementary,lostaglio2022continuous,korzekwa2022optimizing,czartowski2023thermal}, to conclusively investigate the role of non-Markovianity in thermodynamics we may need to resort to an alternative approach.

Let us adopt a practical standpoint where an experimenter can access the system at a discrete set of times. The dynamics is then better described by means of the \textit{ process tensor} framework \cite{pollock2018non}. A desirable feature of this approach is that  the  quantum comb structure \cite{chiribella2009theoretical} of process tensors allows a natural definition of quantum (non-)Markovianity \cite{pollock2018operational}, removing any ambiguity present in other approaches \cite{rivas2014quantum,breuer2016colloquium,de2017dynamics,li2018concepts,milz2019completely,chruscinski2022dynamical}. This makes process tensors useful for understanding memory effects in quantum dynamics \cite{taranto2019quantum,taranto2019structure,figueroa2019almost,milz2020kolmogorov,milz2020when,milz2021genuine,taranto2021non,figueroa2021markovianization,sakuldee2022connecting,capela2022quantum,taranto2023hidden,Berk2021resourcetheoriesof,taranto2024characterising,zambon2024relations,santos2024quantifying} and revisiting problems that had mostly been treated under the Markov hypothesis, such as quantum process tomography \cite{milz2018reconstructing,white2020demonstration,white2022non,white2022characterization,white2023filtering,aloisio2023sampling}, simulation \cite{jorgensen2019exploiting,jorgensen2020discrete,cygorek2022simulation,fowler2022efficient,gribben2022using,cygorek2024sublinear,fux2023tensor}, and thermalization \cite{strasberg2019repeated,figueroa2020equilibration,huang2022fluctuation,huang2023multiple,dowling2023relaxation,dowling2023equilibration}, among others \cite{figueroa2021randomized,figueroa2022towards,berk2023extracting,figueroa2024operational,butler2024optimizing,zambon2024process}.

In this Letter, we establish non-Markovianity of quantum processes as a fundamental resource for work extraction in thermodynamics. Building on recent results for work cost and work distillable from quantum states \cite{horodecki2013fundamental,brandao2013resource,brandao2015second} and channels \cite{navascues2015nonthermal,faist2019thermodynamic,faist2021thermodynamic}, we quantify the net work $W$ extractable from general multitime quantum process tensors. We define a hierarchy of four classes of strategies \cite{cnotchoi}, labeled, respectively, as {\it sequential}, {\it joint}, {\it global}, and {\it comb}, which enable progressively more work to be extracted. Crucially, for a Markovian process, these strategies are all equivalent and the hierarchy collapses. On the other hand, for a general process we identify the three mechanisms of work investment, multitime correlations, and system-environment correlations, through which non-Markovianity strictly enhances work extraction at every step up the hierarchy:
\begin{equation}\label{hierarchy}
W^{\textup{seq}} \underset{\!\!\!\!\!\!\!{\small{\begin{array}{c}\uparrow \\ \textup{work} \\ \textup{investment}\end{array}}}\!\!\!\!\!\!\!}{\leq}
W^{\textup{joint}} \underset{\!\!\!\!\!{\small{\begin{array}{c}\uparrow \\ \textup{multitime} \\ \textup{correlations}\end{array}}}\!\!\!\!\!}{\leq}
W^{\textup{global}} \underset{\!\!\!\!\!\!\!\!\!\!\!\!\!\!\!{\small{\begin{array}{c}\uparrow \\ \textup{system-environment} \\ \textup{correlations}\end{array}}}\!\!\!\!\!\!\!\!\!\!\!\!\!\!\!}{\leq}
W^{\textup{comb}}\,.
\end{equation}
Although these mechanisms may coexist in general, we provide examples in which their action is isolated, validating our definitions and interpretation. We also derive quantitative bounds precisely linking the enhancement provided by each mechanism to the amount of non-Markovianity present in the process. Our results bridge the gap between quantum dynamics and thermodynamics, providing a fundamental characterization of the advantages unlocked by non-Markovian processes in thermodynamic tasks.

\begin{figure*}[t]
\captionsetup[subfloat]{labelformat=empty,captionskip=-24pt}
    \centering
    \subfloat[]{\includegraphics[width=0.85\linewidth]{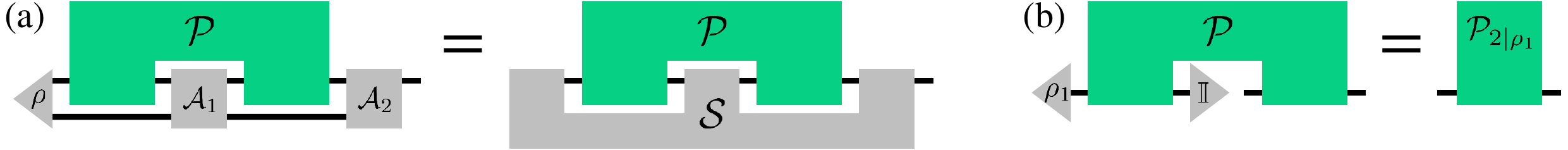}\label{fig:combinedsa}}\subfloat[]{\label{fig:combinedsb}}
    \caption{(a) General multitime quantum process. The experimenter prepares an initial state $\rho$ for the system and then performs the control operations described by the channels $\Acal_1$ and $\Acal_2$. In between operations the system interacts with an uncontrolled environment. The dynamics is described by a process tensor $\Pcal$ mapping the control operations to the final state of the system. The physical constraints on the process imply $\Pcal$ is a quantum comb. Since the control operations may also be correlated by an ancilla, they are also in general described by a comb $\Scal$.
    (b) Channel $\Pcal_{2|\rho_1}$ associated with the second step of the dynamics described by $\Pcal$ once the input state $\rho_1$ of the first step is specified and the output discarded. In general, the channel describing the evolution in any given step of the dynamics will be conditioned on all previous inputs, but never on the subsequent ones since the process is time ordered.}
    \label{fig:combineds}
\end{figure*}

\textit{Setup.}--- We consider the resource-theoretic scenario in which an experimenter can perform \textit{ thermal operations} for free to implement thermodynamic tasks \cite{janzing2000thermodynamic,brandao2013resource,Ng2018,lostaglio2019introductory}. That is, given a bath $B$ with Hamiltonian $H_B$ and temperature $T$ and a system $S$ with (time-independent) Hamiltonian $H_S$ \cite{anottime}, the operations $\Ecal_{TO}$ the experimenter can perform on the state $\rho_S$ of the system are of the form
\begin{equation}
    \label{eq:TO}
    \Ecal_{TO}(\rho_S) = \tr_{B}\qty[U(\rho_S\otimes\gamma_B)U^\dagger],
\end{equation}
where $\gamma_B \Def e^{-H_B/kT}/\tr[e^{-H_B/kT}]$ is the bath thermal state, $k$ is Boltzmann's constant, and $U$ is a unitary operator that acts jointly on system and bath ensuring conservation of the total energy,  $[U,H_S+H_B]=0$.

The goal of the experimenter is to extract, by means of thermal operations, the maximum amount of work from any given nonthermal resource--which could be a state, a channel, or a multitime process--in the asymptotic setting in which $n \rightarrow \infty$ copies of the given resource are available. The work $W(\rho_S)$ asymptotically distillable from a general state $\rho_S$ is known to be given by $kTS(\rho_S||\gamma_S)$, where $S(\rho||\sigma)=\tr[\rho(\ln \rho-\ln\sigma)]$ is the relative entropy between the states $\rho$ and $\sigma$ \cite{horodecki2013fundamental,brandao2013resource,brandao2015second}. Reference \cite{brandao2013resource} also showed that state conversion is asymptotically reversible, implying that the work needed to transform a thermal state into another state $\rho$ is also given by $W(\rho)$. Similarly, work extraction from a nonthermal channel $\Ecal$ was discussed in Ref.~\cite{navascues2015nonthermal}, which showed that the best protocol consists of preparing some optimal state $\rho$ at a cost $W(\rho)$, then using the channel to obtain the state $\Ecal(\rho)$, and finally extracting $W(\Ecal(\rho))$ of work in the asymptotic limit. In this way, the maximum work distillable from a channel $\Ecal$ is given by $W(\Ecal)\Def\max_{\rho}\{W[\Ecal(\rho)]-W(\rho)\}$.

We now take the next step in this direction and analyze work extraction from general multitime quantum processes. Consider that the experimenter can prepare an initial state and then perform a sequence of control operations on the system. In between these operations the system interacts with an uncontrolled environment, giving rise to a multitime open quantum system dynamics. Such a dynamics is described by a process tensor $\Pcal$ mapping the sequence of control operations $\Scal$ to the final state of the system \cite{pollock2018non}. The physical constraints on the process imply that, for a process with an initial state preparation and $n-1$ control operations, $\Pcal$ belongs to the set $\Pbb_n$ of $n$-step \textit{ quantum combs} \cite{chiribella2009theoretical}. Consequently, we have $\Scal\in\Sbb_n$, where $\Sbb_n$ is the set of quantum combs mapped by the elements of $\Pbb_n$ to a final state $\sigma=\Pcal(\Scal)$, as in Fig.~\subref{fig:combinedsa}.

Importantly, the structure of $\Pcal$ is also the same as that of a sequence of quantum channels with memory \cite{kretschmann2005quantum}, so it can be seen as an ordered mapping from a set of input states to a set of output states. In this sense, while the first step of the process is always a proper channel $\Pcal_1$, the subsequent steps may be channels conditioned on the inputs at previous times, as in Fig.~\subref{fig:combinedsb}.
However, if that is not the case and the process consists of a sequence of independent channels, we say the process is \textit{ Markovian}, as no memory is carried along the dynamics \cite{pollock2018operational}. This implies that extracting work from a Markovian process is the same as extracting work from a set of channels, for which one can apply the protocol of Ref.~\cite{navascues2015nonthermal} and the work distillable from the process will simply be the sum of the works distillable from each channel. Therefore, any fundamental difference between extracting work from channels and general processes can only come from non-Markovian effects. In this sense, it is crucial to relate the amount of extractable work to the amount of non-Markovianity of the process.

To quantify the non-Markovianity of any given process $\Pcal\in\Pbb_n$ we employ the following distinguishability measure with respect to the closest Markovian process:
\begin{equation}
\label{eq:nm-def}
    N(\Pcal)\Def \min_{\Qcal\in\Pbb_n^{{M}}}\bar{S}(\Pcal||\Qcal),
\end{equation}
where $\Pbb_n^{{M}}$ is the set of Markovian $n$-step process tensors and
\begin{equation}
    \bar{S}(\Pcal||\Qcal)\Def \max_{\Scal\in\Sbb_n}S[\Pcal(\Scal)||\Qcal(\Scal)]
\end{equation}
is the relative entropy between  process tensors $\Pcal$ and $\Qcal$ \cite{zambon2024process}. This is a fully bona fide measure fulfilling the prescriptions demanded by the resource theory of non-Markovianity \cite{Berk2021resourcetheoriesof}.

Having set the stage with background results and definitions, we now proceed to the hierarchy (\ref{hierarchy}) of work extraction protocols for multitime quantum processes. 
We begin with the simplest one, which makes no use of the non-Markovianity of the process and is thus set as a reference for the other ones.

\textit{Sequential optimization.}--- The first protocol we propose is characterized by sequential optimization of inputs and local extraction of work. As described in Fig.~\subref{fig:combinedsb}, for any process $\Pcal\in\Pbb_n$, the channel $\Pcal_1$ associated with its first step is always well defined. Therefore, for this channel, we could apply the protocol from Ref.~\cite{navascues2015nonthermal} to extract $W(\Pcal_1)$ of work in the first step. This is achieved by preparing an optimal state $\rho_1$ and feeding it to the first step of the process, then extracting work from the first output state $\Pcal_1(\rho_1)$. After using $\rho_1$ as the first input, the second step of the process will be described by a channel $\Pcal_{2|\rho_1}$, for which the procedure could be repeated, leading to the extraction of $W(\Pcal_{2|\rho_1})$ of work. The sequential optimization protocol consists of iterating this procedure until the last step of the process. A rigorous definition for the total work $W^{\textup{seq}}(\Pcal)$ extracted through the sequential optimization protocol is given in the \hyperref[sec:appendix]{Appendix}.

\begin{figure*}[bt]
\captionsetup[subfloat]{captionskip=-6pt}
    \centering
\subfloat[]{\label{fig:joint}\includegraphics[width=0.32\linewidth]{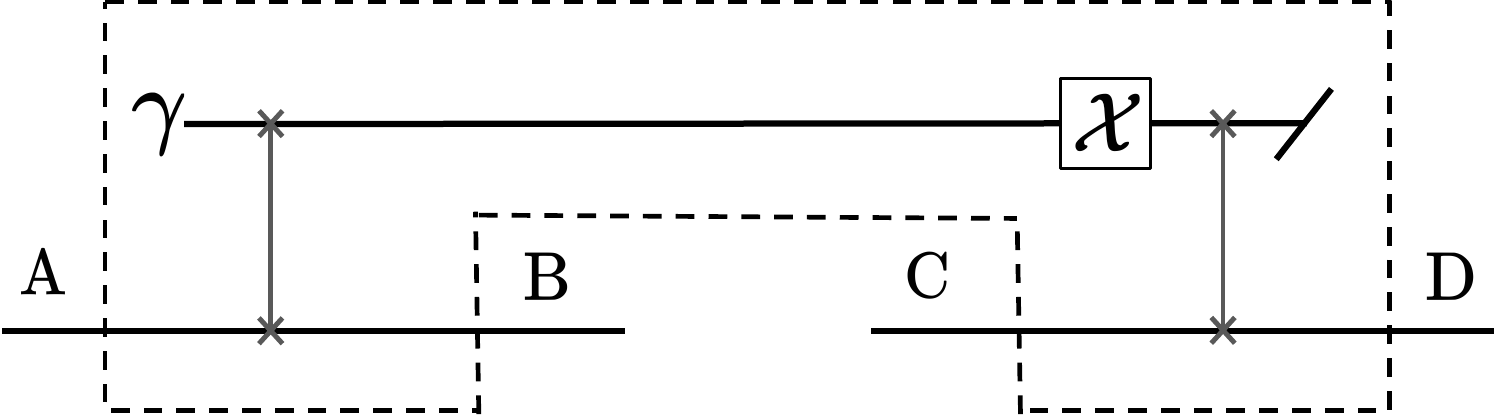}}\quad
\subfloat[]{\label{fig:global}\includegraphics[width=0.32\linewidth]{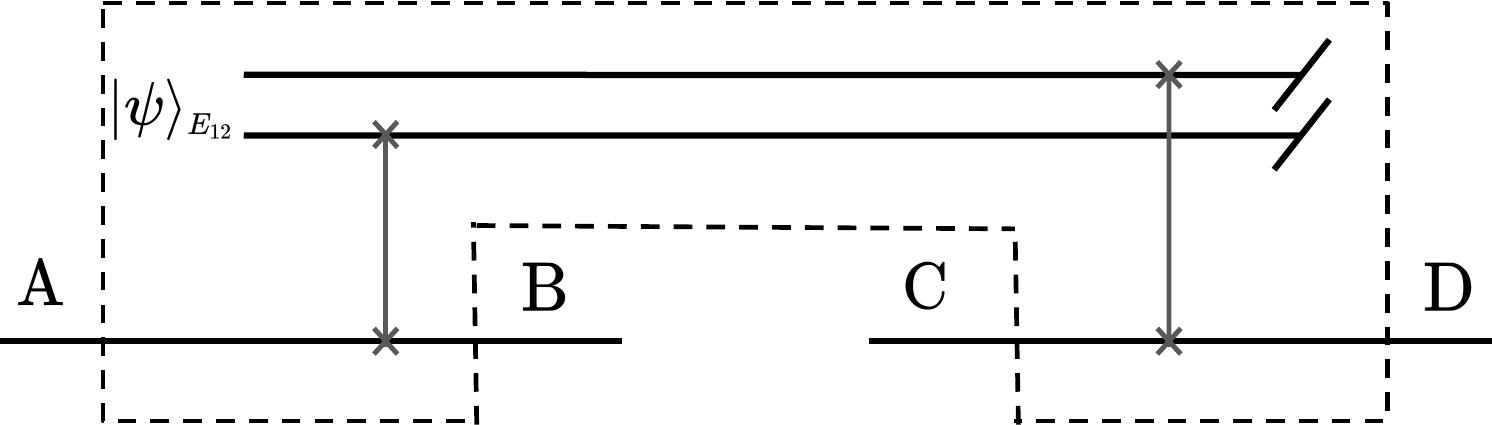}}\quad
\subfloat[]{\label{fig:general}\includegraphics[width=0.32\linewidth]{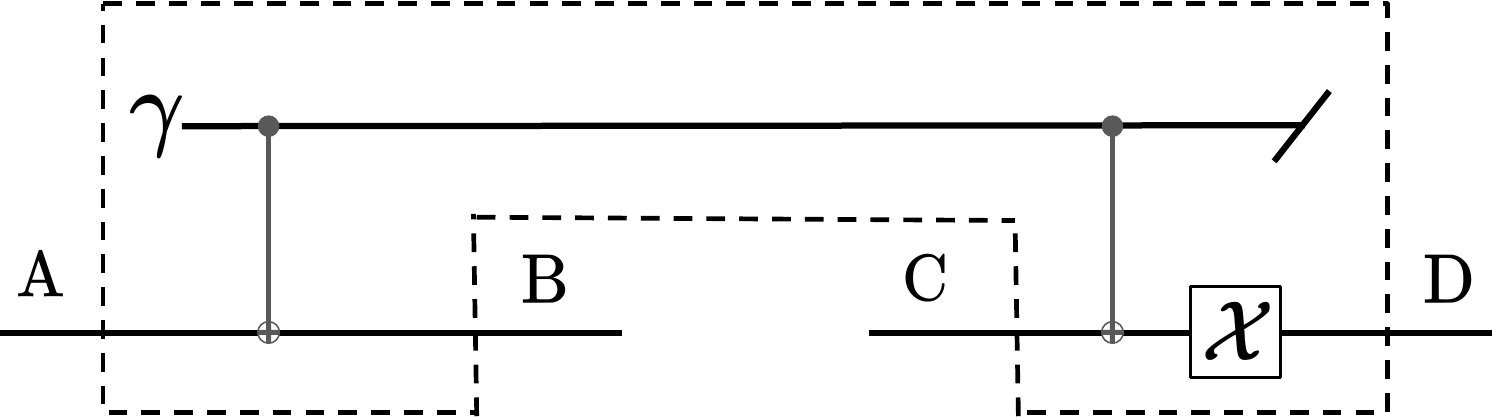}}
\caption{Examples of non-Markovian processes. (a) A process where sequential optimization is not optimal. Both system and  environment are qubits with Hamiltonian $H=E\ket{1}\bra{1}$. The initial state of the environment is thermal and the first step of the process is a SWAP gate between system and environment, while the second step is a NOT gate on the environment, $\Xcal(\bullet)=X\bullet X^\dagger$, $X=\ket{0}\bra{1}+\ket{1}\bra{0}$, followed by another SWAP gate. (b) A process where no work can be extracted through joint optimization. The environment is composed by two qubits $E_1$ and $E_2$ with Hamiltonians $H_{E_1}=H_{E_2}=E\ket{1}\bra{1}$. The system is also a qubit with the same Hamiltonian. The initial state of the environment is $\ket{\psi}_{E_{12}}\propto\ket{00}+e^{-E/2kT}\ket{11}$, such that it is locally thermal despite being globally pure. Each step of the process is a SWAP interaction between the system and a single part of the environment, resembling a collisional model but allowing for non-Markovianity since the ancillas are initially correlated. (c) A process where global extraction is not optimal. Both system and environment are qubits with Hamiltonian $H=E\ket{1}\bra{1}$. The environment is initially thermal at temperature $T$. The first step of the dynamics consists of a controlled-NOT (CNOT) gate with control on the environment and target on the system. In the second step we have the same CNOT gate, but this time followed by a NOT gate operation on the system.}
\end{figure*}

\textit{Joint optimization.}--- While the result of Ref.~\cite{navascues2015nonthermal} implies $W^{\textup{seq}}(\Pcal)$ is the maximum work extractable from Markovian processes, this does not hold in general in the presence of non-Markovianity. Consider, for example, the process $\Pcal$ of Fig.~\subref{fig:joint}. Notice that the channel $\Pcal_1$ has the thermal state $\gamma_S$ as a fixed output, such that no work can be extracted from its output independently of the input. In this case, sequential optimization implies the optimal first input is $\rho_1=\gamma_S$, as it can be prepared for free, yielding $W(\Pcal_1)=0$. This implies the second channel $\Pcal_{2|\rho_1}$ has $\Xcal(\gamma_S)$ as a fixed output. Again, the optimal input is $\rho_2=\gamma_S$, yielding $W^{\textup{seq}}(\Pcal)=kTS[\Xcal(\gamma_S)||\gamma_S]=(1-e^{-E/kT})E$.

It is possible, however, to make use of the non-Markovianity of the process to extract even more work from it. If the experimenter changes the first input to $\rho_1=\ket{0}\bra{0}$, at a cost of $kT\ln(1+e^{-E/kT})$, the second output will be $\ket{1}\bra{1}$, from which one can extract $E+kT\ln(1+e^{-E/kT})$ of work, resulting in a net distilled work equal to $E$. This advantage with respect to the sequential optimization comes from the fact that the non-Markovianity of the process allows the experimenter to spend some work in the first step to extract even more in the second one. This mechanism is what we call \textit{work investment} (WI).

To make full use of work investment, it is necessary to perform a \textit{joint optimization protocol}, in which the inputs of each step of the process are chosen such that the total work distilled is maximized. The work $W^{\textup{joint}}(\Pcal)$ extractable through joint optimization is mathematically defined in the \hyperref[sec:appendix]{Appendix}. Since this construction encompasses the cases where sequential optimization is optimal, it is immediately clear that $W^{\textup{joint}}(\Pcal)$ is always greater than or equal to $W^{\textup{seq}}(\Pcal)$, with the drawback that in practice the joint optimization is much harder to be computed than the sequential one.

Knowing that sequential optimization is optimal when the process is Markovian, we should expect the advantage of work investment to be little if the amount of non-Markovianity in the process is little. To validate such an intuition, we define the additional work extractable from work investment as $\Delta W^{\textup{WI}}(\Pcal)\Def W^{\textup{joint}}(\Pcal)-W^{\textup{seq}}(\Pcal)$, and in Supplemental Material \cite{supp} we prove the following continuity bound.

\begin{theorem}\label{thm:wi}
    For any $n$-step process $\Pcal\in\Pbb_n$ with non-Markovianity $N(\Pcal)$, describing the evolution of a system $S$ with Hamiltonian $H_S$ and in contact with a bath at temperature $T$, it holds that
    \begin{equation}
        \Delta W^{\textup{WI}}(\Pcal)\le kT~\Fcal(H_S,T,n)[N(\Pcal)]^{1/4},
    \end{equation}
    where
    \begin{equation*}
        \Fcal(H_S,T,n)=2^{1/4}\qty[\sqrt{2}\ln 2+S(\Pi_{E_{\textup{max}}}||\gamma_S)](n-1),
    \end{equation*}
    and $\Pi_{E_{\textup{max}}}=\ket{E_{\textup{max}}}\bra{E_{\textup{max}}}$ is the projector onto the most energetic eigenstate of $H_S$.
\end{theorem}

Being a continuity bound, the above inequality is not tight in general \cite{bluhm2023continuity}, but it is important to show that the amount of non-Markovianity quantitatively limits the thermodynamic yield obtainable from an initial work investment. This is of practical relevance, since typical processes have small non-Markovianity if the environment is large  enough \cite{figueroa2019almost}.

\textit{Global optimization.}--- Work investment, however, is not the only mechanism through which non-Markovianity can enhance work extraction. To see this, consider the process of Fig.~\subref{fig:global}. Notice that, independently of the chosen inputs $\rho_1$ and $\rho_2$, the two outputs of the process will locally be thermal states $\gamma_S$, from which no work can be extracted. Nevertheless, if the experimenter has access to an auxiliary quantum system, which we call a \textit{side memory} \cite{bnotancilla}, where the first output of the process can be stored until after the second step, it is possible to perform work extraction from the global output state $\rho_{12}=\ket{\psi}\bra{\psi}_{{12}}$, obtaining $W(\rho_{12})=2kTS(\gamma)>0$. This happens because non-Markovian dynamics may create correlations between outputs at different times, even if the inputs were initially uncorrelated. This mechanism is what we call \textit{multitime correlations} (MTC).

To fully explore the multitime correlations that result from the process, the experimenter must perform a \textit{global optimization protocol}, in which all the outputs are stored until the end and work is distilled from the global final state. The inputs are chosen as to achieve the maximum globally extractable work $W^{\textup{global}}(\Pcal)$. Besides the computational complexity of carrying out such optimizations, the implementation of this protocol is likely to be experimentally challenging in most cases, as the amount of side memory resources scales with $n$. Still, from a theoretical standpoint, it shows a second way through which non-Markovianity influences work extraction. Importantly, the work extractable from multitime correlations $\Delta W^{\textup{MTC}}(\Pcal)\Def W^{\textup{global}}(\Pcal)-W^{\textup{joint}}(\Pcal)$ is directly related to the non-Markovianity of the process through the following result, which is also proven in Supplemental Material \cite{supp}.

\begin{theorem}\label{thm:mtc}
    For any $n$-step process $\Pcal\in\Pbb_n$ with non-Markovianity $N(\Pcal)$, describing the evolution of a system $S$ in contact with a bath at temperature $T$, it holds that
    \begin{equation}
        \Delta W^{\textup{MTC}}(\Pcal)\le kTN(\Pcal).
    \end{equation}
\end{theorem}

Another meaningful way to understand the result above is by defining the maximum work $W_i^{\textup{max}}(\Pcal)$ locally extractable in the $i$-th step of a process $\Pcal$ as $W(\Pcal_{i|\bm{r}})$ maximized over all possible vectors $\bm{r}$ of previous input states. Theorem~\ref{thm:mtc} then implies that,  while for joint optimization we have $W^{\textup{joint}}(\Pcal)\le \sum_{i=1}^n W_i^{\textup{max}}(\Pcal)$,
    for global extraction we get
        \begin{equation}
    W^{\textup{global}}(\Pcal)\le \mbox{$\sum_{i=1}^n W_i^{\textup{max}}(\Pcal) + kTN(\Pcal)$}.
        \end{equation}
This clearly shows how the global protocol  allows for work distillation using not only the local athermalities of the process, but also the multitime correlations arising from its non-Markovianity as extra resource. This is crucial, for example, in the process of Fig.~\subref{fig:global}, in which the $W_i^{\textup{max}}$ are all zero and yet there is nonzero work to be extracted from the correlations. In fact, in Sec. VI of \cite{supp} we show that, for the example of Fig.~\subref{fig:global}, the bound given by Theorem~\ref{thm:mtc} is tight.

\textit{Comb optimization.}--- All three protocols we have discussed so far involve preparing uncorrelated input states and feeding them into each step of the process. However, as shown in Fig.~\subref{fig:combinedsa}, we could in principle have general channels connecting consecutive time steps. To see how this generalization may lead to a further advantage in work extraction, we consider the process $\Pcal$ from Fig.~\subref{fig:general}. By preparing the first input $\rho_1=\ket{0}\bra{0}$ and using an identity channel $\Ical$ to connect the two steps, we obtain the final state $\ket{1}\bra{1}$, yielding a net extracted work of $E$. On the other hand, using a global extraction protocol we strictly have $W^{\textup{global}}(\Pcal)<E$ \cite{supp}. The advantage in this case comes from the fact that, in the first step of the dynamics, system and environment create correlations that, despite not being immediately useful for work extraction, allow for the state $\ket{1}\bra{1}$ to be achievable in the second step, from which a higher amount of work can be distilled. If instead the experimenter stores the first output and prepares a new input, the state $\ket{1}\bra{1}$ is not achievable anymore because system and environment are not initially correlated. Importantly, this advantage cannot be present in a Markovian dynamics as system-environment correlations are lost in between time steps.  For these reasons, the last mechanism through which non-Markovianity powers up   work extraction is named \textit{system-environment correlations} (SEC).

The protocol that allows the experimenter to use all system-environment correlations present in a process consists of performing general quantum channels on the system in between the steps of the dynamics. To achieve all possible system transformations, these channels may be connected by an ancilla. Notice that, as shown in Fig.~\subref{fig:combinedsa}, this is equivalent to performing a comb $\Scal$ on the system, which is the most general mapping from a process tensor to a final state. Therefore, this class of protocols, named \textit{comb optimization}, is the most general possible for extracting work from a multitime quantum process, containing all previously discussed ones.

Note that, for optimizing over combs to obtain the most efficient protocol, we need to use the fact that the work cost of implementing a general nonthermal channel $\Ecal$ is given by $W(\Ecal)$ \cite{faist2019thermodynamic,faist2021thermodynamic}. With this in mind, we can provide a continuity bound to the work gain $\Delta W^{\textup{SEC}}(\Pcal) \Def W^{\textup{comb}}(\Pcal) - W^{\textup{global}}(\Pcal)$ obtainable through system-environment correlations \cite{supp},  where $W^{\textup{comb}}(\Pcal)$ is the work extractable from the process $\Pcal$ using comb optimization, defined in the \hyperref[sec:appendix]{Appendix}.

\begin{theorem}\label{thm:sec}
    For any $n$-step process $\Pcal\in\Pbb_n$ with non-Markovianity $N(\Pcal)$, describing the evolution of a system $S$ with Hamiltonian $H_S$ and in contact with a bath at temperature $T$, it holds that
    \begin{equation}
      \Delta  W^{\textup{SEC}}(\Pcal)\le kT~ \Gcal(H_S,T,n)[N(\Pcal)]^{1/4},
    \end{equation}
    where
    \begin{equation*}
        \Gcal(H_S,T,n)=2^{1/4}\qty[\sqrt{2}\ln 2+(2n-1)S(\Pi_{E_{\textup{max}}}||\gamma_S)],
    \end{equation*}
    and $\Pi_{E_{\textup{max}}}=\ket{E_{\textup{max}}}\bra{E_{\textup{max}}}$ is the projector onto the most energetic eigenstate of $H_S$.
\end{theorem}

\textit{Conclusion.}--- We have established the fundamental role of non-Markovianity in thermodynamic work extraction from general quantum processes. We defined a hierarchy of four classes of extraction protocols [Eq.~(\ref{hierarchy})] which led to the identification of three mechanisms through which non-Markovianity enhances work extraction: work investment, multitime correlations, and system-environment correlations. We presented examples and bounds that show how the advantage these mechanisms provide disappears when the process has a vanishing degree of non-Markovianity. The combined effect of these mechanisms [Thms~\ref{thm:wi}--\ref{thm:sec}] leads to a remarkable characterization  of the work extractable from an $n$-step quantum process $\Pcal$ under the most general comb optimization,
\begin{equation}\label{eq:thebound}
W^{\textup{comb}}(\Pcal) = W^{\textup{seq}}(\Pcal) + \Delta W^N(\Pcal)\,,
\end{equation}
where $W^{\textup{seq}}$ is the best that one can do in the Markovian case, while $\Delta W^{N} \Def \Delta  W^{\textup{WI}}+\Delta  W^{\textup{MTC}}+\Delta  W^{\textup{SEC}}$ captures the additional work obtainable by non-Markovian resources, which is  bound by a {\it monotonic} function of the non-Markovianity degree $N(\Pcal)$ of the process, $\Delta W^{N} (\Pcal) \leq kT \big[N(\Pcal) + \big((3 n-2) S(\Pi_{E_{\textup{max}}}||\gamma_S)+\sqrt{2}\ln2\ n\big)\big(2 N(\Pcal)\big)^{1/4}\big]$.

Our work establishes a new qualitative and quantitative framework to investigate non-Markovian effects in quantum thermodynamics, which may be seen to complement that of Refs.~\cite{lostaglio2018elementary,lostaglio2022continuous,korzekwa2022optimizing,czartowski2023thermal} in the context of athermality resource theories. Moreover, our results extend those of Ref.~\cite{bylicka2016thermodynamic} and help understand the thermodynamic advantages shown in Refs. \cite{strasberg2016nonequilibrium,thomas2018thermodynamics,pezzutto2019out,abiuso2019non,kamin2020non,bhattacharya2020thermodynamic,debiossac2020thermodynamics,abah2020implications,camati2020employing,shirai2021non,halpern2020fundamental,spaventa2022capacity,ptaszy2022non,cheong2023effects,picatoste2024dynamically,mohammadi2024quantum} on a more fundamental level. The bounds we proved here may also be employed to define thermodynamic witnesses of non-Markovianity, and by splitting the temporal correlations into classical and quantum \cite{Zurek2003,Deficit2002,Liuzzo2016thermodynamics,ABC,wilde2013quantum} one could obtain witnesses of purely quantum memory for general dynamics \cite{giarmatzi2021witnessing,taranto2024characterising}. We expect our results to stimulate further investigation on the interplay between non-Markovianity and thermodynamics, going beyond the asymptotic regime considered here, to get more experimentally feasible estimates of the work cost and yield of implementing quantum processes with finite resources. This may lead to a deeper comprehension of the energetics of quantum systems \cite{QEI}, potentially inspiring optimal designs for sustainable near-term quantum technologies.

\textit{Acknowledgements}--- G.~Z. acknowledges support by the S{\~a}o Paulo State Foundation (FAPESP) under Grants No. 2022/00993-9 and No. 2023/04625-7. G.~A. acknowledges support by the UK Research and Innovation (UKRI) under BBSRC Grant No. BB/X004317/1 and EPSRC Grant No. EP/X010929/1.

\appendix
\renewcommand{\thedefinition}{A\arabic{definition}}
\section*{End Matter}\label{sec:appendix}

\textit{Definitions.}--- We denote by $\Pcal_{i|\bm{r}}$ the quantum channel describing the $i$-th step of the process $\Pcal\in\Pbb_n$, given the inputs of previous steps were the first $i-1$ entries of the vector $\bm{r}=(\rho_1,\cdots,\rho_{i-1},\cdots,\rho_n)$. Then, let $\rho_{i|\bm{r}}^{\textup{max}}$ be the state maximizing the work extractable from the channel $\Pcal_{i|\bm{r}}$, that is,
\begin{equation}
    \rho_{i|\bm{r}}^{\textup{max}}\Def \argmax_\rho\qty[W(\Pcal_{i|\bm{r}}(\rho))-W(\rho)].
\end{equation}
This allows for the following definition.
\begin{definition}
Let $\bm{r}^{\textup{seq}}$ be the vector of states sequentially optimizing the work extraction from a process $\Pcal\in\Pbb_n$, that is, $\bm{r}^{\textup{seq}}_1=\rho_1^{\textup{max}}$ and $\bm{r}^{\textup{seq}}_i=\rho_{i|(\bm{r}^{\textup{seq}}_{1},\cdots,\bm{r}^{\textup{seq}}_{i-1})}^{\textup{max}}$. The total work $W^{\textup{seq}}(\Pcal)$ extractable from a process $\Pcal\in\Pbb_n$ through the sequential optimization protocol is given by
\begin{equation}
    W^{\textup{seq}}(\Pcal)\Def \sum_{i=1}^n W(\Pcal_{i|\bm{r}^{\textup{seq}}}).
\end{equation}
\end{definition}

The joint optimization protocol is given by relaxing the sequential optimization to a joint one, as follows.
\begin{definition}
The total work $W^{\textup{joint}}(\Pcal)$ extractable from a process $\Pcal\in\Pbb_n$ through the joint optimization protocol is given by
\begin{equation}
    W^{\textup{joint}}(\Pcal)\Def \max_{\bm{r}} \sum_{i=1}^n W(\Pcal_{i|\bm{r}}),
\end{equation}
where the maximization is over all possible vectors $\bm{r}$ of inputs.
\end{definition}

For the global optimization, one can define a class of combs that implement it. Let $\Sbb^{\textup{global}}_n\subset\Sbb_n$ be the set of combs consisting of inputting state $\bm{r}_i=\rho_i$ to the $i$-th step and storing all the outputs until the end. Any such comb $\Scal_{\bm{r}}$ is uniquely defined by the vector $\bm{r}$ of states it inputs to the process. The implementation cost $W(\Scal_{\bm{r}})$ is simply $\sum_{i=1}^n W(\rho_i)$ and the work extractable in the end is $W[\Pcal(\Scal_{\bm{r}})]$, leading to the following definition.

\begin{definition}
The total work $W^{\textup{global}}(\Pcal)$ extractable from a process $\Pcal\in\Pbb_n$ through the global optimization protocol is given by
\begin{equation}
    W^{\textup{global}}(\Pcal)\Def \max_{\Scal_{\bm{r}}\in\Sbb^{\textup{global}}_n} \{W[\Pcal(\Scal_{\bm{r}})]-W(\Scal_{\bm{r}})\}.
\end{equation}
\end{definition}

Finally, to define comb optimization we just need to clarify the cost of implementing general combs. Unlike the case of states and channels, it is not known whether the work cost of implementing a comb is asymptotically equal to the work extractable from it. Since any comb $\Scal\in\Sbb_n$ can be dilated as the action of $n$ channels $\Ecal_i$ acting both on the system and an ancilla initially in the state $\sigma_A$, we take the cost $W(\Scal)$ to be simply $W(\sigma_A)+\sum_{i=1}^n W(\Ecal_i)$ minimized over all possible dilations of $\Scal$. This leads us to the last definition.

\begin{definition}
The total work $W^{\textup{comb}}(\Pcal)$ extractable from a process $\Pcal\in\Pbb_n$ through comb optimization is given by
\begin{equation}
    W^{\textup{comb}}(\Pcal)\Def \max_{\Scal\in\Sbb_n} \{W[\Pcal(\Scal)]-W(\Scal)\}.
\end{equation}
\end{definition}


\begin{thebibliography}{102}
\bibitem [{bno()}]{bnotancilla}%
  \BibitemOpen
  \href@noop {} {}\bibinfo {note} {This auxiliary system is considered ideal for internal consistency of the framework. Note that the system of interest is anything the experimenter has access to that also interacts with an uncontrolled environment, represented by the black horizontal line on the top of Fig. \ref{fig:combinedsa}. On the other hand, the ancilla is, by definition, only controlled by the experimenter, as represented by the black horizontal line on the bottom of Fig. \ref{fig:combinedsa}.}\BibitemShut {Stop}%
\end{thebibliography}

%

\onecolumngrid
\section*{SUPPLEMENTAL MATERIAL}

\renewcommand{\theequation}{S\arabic{equation}}
\renewcommand{\thefigure}{S\arabic{figure}}
\renewcommand{\thelemma}{S\arabic{lemma}}
\renewcommand\thesubfigure{(\alph{subfigure})}

\section{Proof of Theorem 1}

We begin by stating the following result from Ref.~\cite{bluhm2023continuity}, which is then used to prove Lemma \ref{lem:cont}.

\begin{lemma}\label{lem:continuity}\cite{bluhm2023continuity}\footnote{There is an error in the published Theorem 5.18 of Ref.~\cite{bluhm2023continuity}, from which we derive Lemma \ref{lem:continuity}. Our lemma is not affected by this error, as we are considering the particular case $\sigma_1=\sigma_2\Def\sigma$. For a corrected version of Ref.~\cite{bluhm2023continuity}, see its latest arXiv version.}
    For states $\rho_i$ and $\sigma$, with $ker~\sigma\subseteq ker~\rho_i$ and $\Tilde{m}\rho_i<\sigma$, $i\in\{1,2\}$, it holds
    \begin{equation}
        |S(\rho_1||\sigma)-S(\rho_2||\sigma)|\le \qty(\ln 2+\frac{\ln \Tilde{m}^{-1}}{\sqrt{2}})||\rho_1-\rho_2||^{1/2},
    \end{equation}
    where $||\rho_1-\rho_2||=\tr[|\rho_1-\rho_2|]$ is the trace distance between $\rho_1$ and $\rho_2$.
\end{lemma}

Notice that to satisfy the condition $\Tilde{m}\rho_i<\sigma$ above it is sufficient to choose $\Tilde{m}$ to be $\sigma_{\textup{min}}$, the minimal nonzero eigenvalue of $\sigma$. Also, by applying the trace distance triangle inequality \cite{wilde2013quantum}, $||\rho_1-\rho_2||\le||\rho_1-\tau||+||\rho_2-\tau||$, and Pinsker inequality \cite{wilde2013quantum}, $S(\rho||\tau)\le ||\rho-\tau||^{2}/2$, we get the following lemma.

\begin{lemma}
\label{lem:cont}
For states $\rho_i$, $\sigma$, and $\tau$, with $ker~\sigma\subseteq ker~\rho_i$, $i\in\{1,2\}$, it holds
    \begin{equation}
    \begin{aligned}
        |S(\rho_1||\sigma)-S(\rho_2||\sigma)|&\le 2^{1/4}\qty(\ln 2+\frac{\ln\Tilde{m}^{-1}}{\sqrt{2}})\qty(\sqrt{S(\rho_1||\tau)}+\sqrt{S(\rho_2||\tau)})^{1/2}.
    \end{aligned}
\end{equation}
\end{lemma}

Now, to prove Theorem 1, we start by showing that the non-Markovianity of the process upper bounds how different the outputs of a given step can be, given the same input at that step but different inputs at previous steps. First, for a process $\Pcal\in\Pbb_n$ let $\Vcal\in\Pbb^{M}_n$ be the closest Markovian process with respect to the measure $N$, that is,
\begin{equation}
    N(\Pcal)= \max_{\Scal\in\Sbb_n}S(\Pcal(\Scal)||\Vcal(\Scal)).
\end{equation}
By restricting the optimization to the subset $\Sbb^{\textup{global}}_n$, we obtain
\begin{align}
    N(\Pcal)&\ge \max_{\Scal_{\bm{r}}\in\Sbb^{\textup{global}}_n}S(\Pcal(\Scal_{\bm{r}})||\Vcal(\Scal_{\bm{r}})) \nonumber \\
    &= \max_{\bm{r}} \sum_{i=1}^n S(\Pcal_{i|\bm{r}}(\rho_i)||\Vcal_i(\rho_i)) + I(1:\cdots:n)_{\Pcal(\Scal_{\bm{r}})},\label{eq:nm-corr}
\end{align}
where we used that $\Vcal(\Scal_{\bm{r}})=\bigotimes_{i=1}^n \Vcal_i(\rho_i)$, since $\Vcal$ is Markovian, and also the property $S(\rho_{12}||\sigma_1\otimes\sigma_2)=S(\rho_{1}||\sigma_1)+S(\rho_{2}||\sigma_2)+I(1:2)_{\rho_{12}}$, where $I(1:2)_{\rho_{12}}=S(\rho_{12}||\rho_1\otimes\rho_2)$ is the mutual information between the subsystems 1 and 2 in the state $\rho_{12}$ \cite{wilde2013quantum}. Since the relative entropy is non-negative, we can drop most terms in the above inequality and it will still hold,
\begin{equation}
\label{eq:nm-bound}
    N(\Pcal)\ge S(\Pcal_{i|\bm{r}}(\rho_i)||\Vcal_i(\rho_i)),
\end{equation}
for all $1\le i\le n$ and $\bm{r}$.

Next, we write the work extracted through investment as $\Delta W^{\textup{WI}}(\Pcal)=\sum_{i=1}^n\qty[W(\Pcal_{i|\bm{r}^{\textup{joint}}})-W(\Pcal_{i|\bm{r}^{\textup{seq}}})]$, where $\bm{r}^{\textup{joint}}$ is the vector of inputs maximizing joint work extraction. The definition of the sequential extraction protocol implies the first term of this sum is smaller or equal than zero, so we can drop it and obtain
\begin{equation}
\label{eq:inv-sum}
    \Delta W^{\textup{WI}}(\Pcal)\le\sum_{i=2}^n\qty[W(\Pcal_{i|\bm{r}^{\textup{joint}}})-W(\Pcal_{i|\bm{r}^{\textup{seq}}})].
\end{equation}
Then, notice that,
\begin{align}
    \frac{1}{kT}W(\Pcal_{i|\bm{r}^{\textup{seq}}})&=S(\Pcal_{i|\bm{r}^{\textup{seq}}}(\rho^{\textup{seq}}_i)||\gamma)-S(\rho^{\textup{seq}}_i||\gamma)\nonumber \\
    &\ge S(\Pcal_{i|\bm{r}^{\textup{seq}}}(\rho^{\textup{joint}}_i)||\gamma)-S(\rho^{\textup{joint}}_i||\gamma),
\end{align}
where the inequality follows from the definition of $\bm{r}^{\textup{seq}}$. This implies
\begin{align}
    \frac{1}{kT}\qty[W(\Pcal_{i|\bm{r}^{\textup{joint}}})-W(\Pcal_{i|\bm{r}^{\textup{seq}}})]&= \qty[S(\Pcal_{i|\bm{r}^{\textup{joint}}}(\rho^{\textup{joint}}_i)||\gamma)-S(\rho^{\textup{joint}}_i||\gamma)] - \qty[S(\Pcal_{i|\bm{r}^{\textup{seq}}}(\rho^{\textup{seq}}_i)||\gamma)-S(\rho^{\textup{seq}}_i||\gamma)] \nonumber\\
    &\le S(\Pcal_{i|\bm{r}^{\textup{joint}}}(\rho^{\textup{joint}}_i)||\gamma)-S(\Pcal_{i|\bm{r}^{\textup{seq}}}(\rho^{\textup{joint}}_i)||\gamma),
\end{align}
to which we can apply Lemma \ref{lem:cont} with $\rho_1=\Pcal_{i|\bm{r}^{\textup{joint}}}(\rho^{\textup{joint}}_i)\Def \rho^{j}_i$, $\rho_2=\Pcal_{i|\bm{r}^{\textup{seq}}}(\rho^{\textup{joint}}_i)\Def \rho^{s}_i$, $\sigma=\gamma$, and $\tau=\Vcal_i(\rho^{\textup{joint}}_i)\Def \tau_i$, yielding
\begin{align}
    \frac{1}{kT}\qty[W(\Pcal_{i|\bm{r}^{\textup{joint}}})-W(\Pcal_{i|\bm{r}^{\textup{seq}}})]&\le 2^{1/4}\qty(\ln 2+\frac{\ln \Tilde{m}^{-1}}{\sqrt{2}})\qty(\sqrt{S(\rho^{j}_i||\tau_i)}+\sqrt{S(\rho^{s}_i||\tau_i)})^{1/2}.
\end{align}
Then, from Eq. \eqref{eq:nm-bound} we have $S(\rho^{j}_i||\tau_i)\le N(\Pcal)$ and $S(\rho^{s}_i||\tau_i)\le N(\Pcal)$, implying
\begin{align}
    \frac{1}{kT}\qty[W(\Pcal_{i|\bm{r}^{\textup{joint}}})-W(\Pcal_{i|\bm{r}^{\textup{seq}}})]&\le 2^{1/4}\qty(\sqrt{2} \ln 2+\ln\gamma^{-1}_{\textup{min}})\qty[N(\Pcal)]^{1/4},
\end{align}
which can be used in Eq. \eqref{eq:inv-sum}, where we just multiply by the number of terms in the sum, as there is no dependence in $i$,
\begin{align}
    \frac{1}{kT}\Delta W^{\textup{WI}}(\Pcal)\le (n-1)2^{1/4}\qty(\sqrt{2} \ln 2+\ln\gamma^{-1}_{\textup{min}})\qty[N(\Pcal)]^{1/4}.
\end{align}

Finally, given $\gamma=e^{-H_S/kT}/Z$ and $\Pi_{E_{\textup{max}}}=\ket{E_{\textup{max}}}\bra{E_{\textup{max}}}$ is the projector onto the most energetic eigenstate of $H_S$, we get
\begin{align}
    S(\Pi_{E_{\textup{max}}}||\gamma)&= \tr[\Pi_{E_{\textup{max}}}\ln\Pi_{E_{\textup{max}}}]- \tr[\Pi_{E_{\textup{max}}}\ln\gamma] \nonumber\\
    &= -\bra{E_{\textup{max}}}\qty(\sum_i \ln\frac{e^{-E_i/kT}}{Z}\ket{E_{i}}\bra{E_{i}} )\ket{E_{\textup{max}}}\nonumber\\
    &= -\ln \frac{e^{-E_{\textup{max}}/kT}}{Z}\nonumber\\
    &= \ln\gamma^{-1}_{\textup{min}},
\end{align}
yielding
\begin{equation}
    \Delta W^{\textup{WI}}(\Pcal)\le kT 2^{1/4}\qty(\sqrt{2} \ln 2+S(\Pi_{E_{\textup{max}}}||\gamma))(n-1)\qty[N(\Pcal)]^{1/4},
\end{equation}
as stated in Theorem 1.

\section{Proof of Theorem 2}

For a process $\Pcal\in\Pbb_n$, let $\Scal_{\bm{r}^{\textup{global}}}\in\Sbb^{\textup{global}}_n$ be the comb maximizing global work extraction, which may be written as
\begin{align}
    W^{\textup{global}}(\Pcal)&=W(\Pcal(\Scal_{\bm{r}^{\textup{global}}}))-W(\Scal_{\bm{r}^{\textup{global}}}) \nonumber\\
    &= \sum_{i=1}^n W(\Pcal_{i|\bm{r}^{\textup{global}}}) + kT I(1:\cdots:n)_{\Pcal(\Scal_{\bm{r}^{\textup{global}}})},
\end{align}
while for joint extraction,
\begin{align}
    W^{\textup{joint}}(\Pcal)&= \max_{\bm{r}} \sum_{i=1}^n W(\Pcal_{i|\bm{r}})\nonumber\\
    &\ge \sum_{i=1}^n W(\Pcal_{i|\bm{r}^{\textup{global}}}),
\end{align}
as $\bm{r}^{\textup{global}}$ does not maximize joint work extraction in general. By combining these two expressions we get
\begin{align}
    W^{\textup{global}}(\Pcal)-W^{\textup{joint}}(\Pcal)&\le kT I(1:\cdots:n)_{\Pcal(\Scal_{\bm{r}^{\textup{global}}})},
\end{align}
and from Eq. \eqref{eq:nm-corr} we have
\begin{align}
\label{eq:N-I}
    N(\Pcal)&\ge\max_{\bm{r}} \sum_{i=1}^n S(\Pcal_{i|\bm{r}}(\rho_i)||\Vcal_i(\rho_i)) + I(1:\cdots:n)_{\Pcal(\Scal_{\bm{r}})}\nonumber\\
    &\ge\max_{\bm{r}} I(1:\cdots:n)_{\Pcal(\Scal_{\bm{r}})}\nonumber\\
    &\ge I(1:\cdots:n)_{\Pcal(\Scal_{\bm{r}^{\textup{global}}})},
\end{align}
finally yielding
\begin{equation}
    W^{\textup{global}}(\Pcal)-W^{\textup{joint}}(\Pcal)\le kT N(\Pcal),
\end{equation}
thus proving Theorem 2.

\section{Example of advantage from system-environment correlations}

In the example of Fig.~2(c), for general inputs $\rho_1$ and $\rho_2$, the final global output state will be given by $\rho^{\prime}_{12}=\gamma_0 \rho_1\otimes\rho_2+\gamma_1 \tilde{\rho}_1\otimes\tilde{\rho}_2$, where $\gamma_i=\bra{i}\gamma\ket{i}$, $i\in\{0,1\}$, and $\tilde{\rho}_j=X\rho_jX$, $j\in\{1,2\}$. Since the quantum relative entropy is both convex and additive under tensor product, we have $S(\rho^{\prime}_{12}||\gamma)\le \gamma_0\qty[S(\rho_1||\gamma)+S(\rho_1||\gamma)]+ \gamma_1\qty[S(\tilde{\rho}_1||\gamma)+S(\tilde{\rho}_1||\gamma)]$, implying
\begin{align}
    W^{\textup{global}} &= \max_{\rho_1,\rho_2}~ W(\rho^{\prime}_{12})-W(\rho_1)-W(\rho_2)\nonumber\\
    &\le \max_{\rho_1,\rho_2}~\gamma_0\qty[W(\rho_1)+W(\rho_2)]+ \gamma_1\qty[W(\tilde{\rho}_1)+W(\tilde{\rho}_2)]-W(\rho_1)-W(\rho_2)\nonumber\\
    &= \gamma_1 \qty{\max_{\rho_1}\qty[W(\tilde{\rho}_1)-W(\rho_1)]+\max_{\rho_2}\qty[W(\tilde{\rho}_2)-W(\rho_2)]}\nonumber\\
    &= 2\gamma_1 W(\Xcal),\label{eq:max}
\end{align}
where we also used $\gamma_0-1=-\gamma_1$, then the definition for the maximum work $W(\Xcal)$ distillable from the unitary bit-flip channel $\Xcal(\bullet)=X\bullet X$. Since $W(\rho)=\tr[\rho H]-kTS(\rho)$ and $S(\rho)=S(\tilde{\rho})$, for $H=E\ket{1}\bra{1}$ we have $W(\Xcal)=\max_\rho \tr[(\tilde{\rho}-\rho) H]=E \max_\rho\bra{1}(\tilde{\rho}-\rho)\ket{1}=E$, maximized by $\rho=\ket{0}\bra{0}$. However, since $\gamma_1=e^{-E/kT}/(1+e^{-E/kT})<1/2$ for finite temperature $T$ and nonzero energy gap $E$, Eq. \eqref{eq:max} yields $W^{\textup{global}}<E$, as stated in the main text.

\section{Proof of Theorem 3}

\begin{figure*}[t]
\captionsetup[subfloat]{labelformat=empty,captionskip=-24pt}
    \centering
    \subfloat[]{\includegraphics[width=0.75\linewidth]{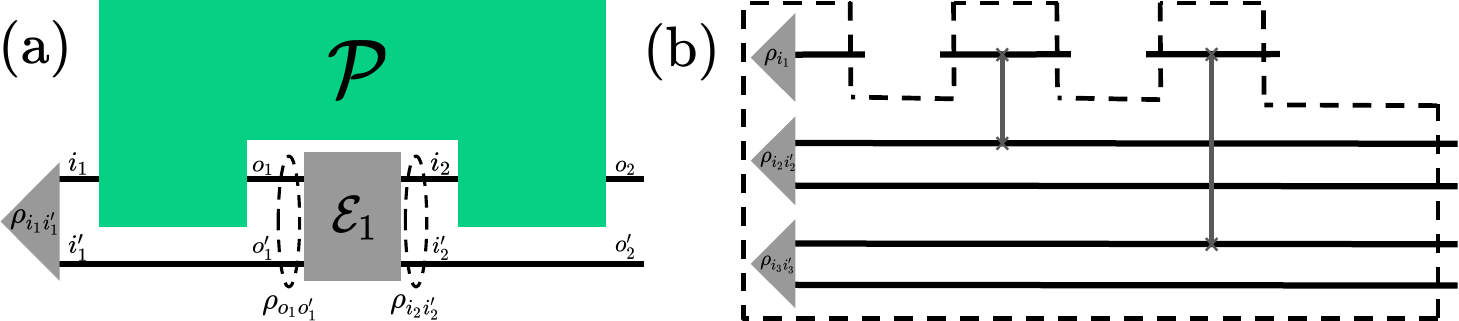}\label{fig:indexes}}\subfloat[]{\label{fig:s2}}
    \caption{(a) Hilbert space indexing for a comb consisting of initial system-ancilla state $\rho_{i_1i^{\prime}_1}$ and a channel $\Ecal_1$, acting on a two-step process tensor $\Pcal\in\Pbb_2$. The state $\rho_{o_1o^{\prime}_1}$ is the output of the first step of the dynamics and also the input of channel $\Ecal_1$. Similarly, the state $\rho_{i_2i^{\prime}_2}$ is the output of channel $\Ecal_1$ and also the input of the second step of the dynamics.
    (b) Example of a comb $\Scal_2$. In this case we have $\Scal_2\in\Sbb_3$, that is, the comb $\Scal_2$ is mapped by 3-step process tensors $\Pcal\in\Pbb_3$ to a final state $\Pcal(\Scal_2)\in\Dcal(\Hcal^{\otimes 5}_S)$.}
    \label{fig:combineds}
\end{figure*}

For any process $\Pcal\in\Pbb_n$, let $\Scal_c\in\Sbb_n$ be the comb maximizing comb work extraction, with a minimal cost dilation given by an initial system-ancilla state $\rho_{i_1i^{\prime}_1}$ and $n-1$ channels $\Ecal_j:\Dcal(\Hcal_{o_j}\otimes\Hcal_{o^{\prime}_j})\to\Dcal(\Hcal_{i_{j+1}}\otimes\Hcal_{i^{\prime}_{j+1}})$, where the indexes $i_j$ and $o_j$ indicate if a given Hilbert space corresponds to an input ($i$) or output ($o$) of the $j$-th step of process $\Pcal$, while the primed index describes the space of the ancilla at that time. This way, the input of the $j$-th step of the dynamics (which is also the output of channel $\Ecal_{j-1}$) is $\rho_{i_ji^{\prime}_j}$, and the output of the $j$-th step of the dynamics (which is also the input of channel $\Ecal_j$) is $\rho_{o_jo^{\prime}_j}$, as shown in Fig.~\ref{fig:indexes}. Let $\bm{r}_c$ be the vector of inputs of $\Pcal$, without ancilla, when $\Pcal$ acts on $\Scal_c$, that is, $\bm{r}_c=(\rho_{i_1},\cdots,\rho_{i_n})$. Clearly,
\begin{align}
    W^{\textup{global}}(\Pcal)&=\max_{\bm{r}}\qty[W(\Pcal(\Scal_{\bm{r}}))-W(\Scal_{\bm{r}})]\nonumber\\
    &\ge W(\Pcal(\Scal_{\bm{r}_c}))-W(\Scal_{\bm{r}_c})\nonumber\\
    &= W(\Pcal(\Scal_{\bm{r}_c}))-\sum_{j=1}^n W(\rho_{i_j}),\label{eq:sum-global}
\end{align}
and also,
\begin{align}
    W^{\textup{comb}}(\Pcal)&=W(\Pcal(\Scal_c))-W(\Scal_c)\nonumber\\
    &= W(\Pcal(\Scal_c))-W(\rho_{i_1i^{\prime}_1})-\sum_{j=1}^{n-1}W(\Ecal_j)\nonumber\\
    &\le W(\Pcal(\Scal_c))-W(\rho_{i_1i^{\prime}_1})-\sum_{j=1}^{n-1}\qty[W(\rho_{i_{j+1}i^{\prime}_{j+1}})-W(\rho_{o_jo^{\prime}_j})]\nonumber\\
    &= W(\Pcal(\Scal_c)) + \sum_{j=1}^{n-1}W(\rho_{o_jo^{\prime}_j}) -\sum_{j=1}^{n}W(\rho_{i_{j}i^{\prime}_{j}}),\label{eq:sum-comb}
\end{align}
where we used
\begin{align}
    W(\Ecal_j)&=\max_\sigma \qty[W(\Ecal_j(\sigma))-W(\sigma)]\nonumber\\
    &\ge W(\Ecal_j(\rho_{o_jo^{\prime}_j}))-W(\rho_{o_jo^{\prime}_j})\nonumber\\
    &=W(\rho_{i_{j+1}i^{\prime}_{j+1}})-W(\rho_{o_jo^{\prime}_j}).
\end{align}

Combining Eqs. \eqref{eq:sum-global} and \eqref{eq:sum-comb} we bound $\Delta W^{\textup{SEC}}(\Pcal)=W^{\textup{comb}}(\Pcal)-W^{\textup{global}}(\Pcal)$ as
\begin{align}
    \Delta W^{\textup{SEC}}(\Pcal)&\le \qty[W(\Pcal(\Scal_c)) + \sum_{j=1}^{n-1}W(\rho_{o_jo^{\prime}_j})] -\qty[W(\Pcal(\Scal_{\bm{r}_c}))+\sum_{j=1}^{n}\qty[W(\rho_{i_{j}i^{\prime}_{j}})-W(\rho_{i_j})]]\nonumber\\
    &\le \qty[W(\Pcal(\Scal_c)) + \sum_{j=1}^{n-1}W(\rho_{o_jo^{\prime}_j})] -\qty[W(\Pcal(\Scal_{\bm{r}_c}))+\sum_{j=1}^{n}\qty[W(\rho_{i^{\prime}_{j}})+kTI(o_{j}:o^{\prime}_{j})_{\Pcal_{j|\bm{r}_c}(\rho_{i_ji^{\prime}_j})}]],\label{eq:brackets1}
\end{align}
where the second inequality comes from
\begin{align}
    W(\rho_{i_{j}i^{\prime}_{j}})-W(\rho_{i_j})&=W(\rho_{i^{\prime}_{j}})+kTI(i_j:i^{\prime}_j)_{\rho_{i_ji^{\prime}_j}}\\
    &\ge W(\rho_{i^{\prime}_{j}})+kTI(o_{j}:o^{\prime}_{j})_{\Pcal_{j|\bm{r}_c}(\rho_{i_ji^{\prime}_j})},
\end{align}
as the channel $\Pcal_{j|\bm{r}_c}$ acts locally on $i_j$. Equation \eqref{eq:brackets1} may be further simplified by noticing that the term $W(\rho_{i^{\prime}_{1}})+kTI(o_{1}:o^{\prime}_{1})_{\Pcal_{1|\bm{r}_c}(\rho_{i_1i^{\prime}_1})}$, from the sum of the second square bracket, also appears inside $W(\rho_{o_1o^{\prime}_1})$, thus canceling out, leading to
\begin{align}
    \Delta W^{\textup{SEC}}(\Pcal)&\le \qty[W(\Pcal(\Scal_c)) + W(\rho_{o_1})+\sum_{j=2}^{n-1}W(\rho_{o_jo^{\prime}_j})] -\qty[W(\Pcal(\Scal_{\bm{r}_c}))+\sum_{j=2}^{n}\qty[W(\rho_{i^{\prime}_{j}})+kTI(o_{j}:o^{\prime}_{j})_{\Pcal_{j|\bm{r}_c}(\rho_{i_ji^{\prime}_j})}]].\label{eq:brackets2}
\end{align}

Now, let $\Scal_1\in\Sbb_n$ be the comb satisfying $\Qcal(\Scal_1)=\Qcal(\Scal_c)\otimes\rho_{o_1}\bigotimes_{j=2}^{n-1}\rho_{o_jo^{\prime}_j},~\forall \Qcal\in\Pbb_n$, that is, $\Scal_1$ implements the comb $\Scal_c$ and also deterministically outputs the states $\rho_{o_1}$ and $\rho_{o_jo^{\prime}_j}$, for $2\le j\le n-1$. Consider also the comb $\Scal_2\in\Sbb_n$ that prepares the states $\rho_{i_1}$ and $\rho_{i_jo^{\prime}_i}$, for $2\le j\le n-1$, feeds the non-primed half into the $j$-th step of the process, and outputs all the primed halves as well as the output of each step, as shown in Fig.~\ref{fig:s2}. By construction, we can rewrite Eq. \eqref{eq:brackets2} as
\begin{align}
    \Delta W^{\textup{SEC}}(\Pcal)&\le W(\Pcal(\Scal_1))-W(\Pcal(\Scal_2))\nonumber\\
    &= kT\qty[S(\Pcal(\Scal_1)||\gamma^{\otimes 2n-1})-S(\Pcal(\Scal_1)||\gamma^{\otimes 2n-1})],
\end{align}
as the outputs of combs $\Scal_1$ and $\Scal_2$ consist of $2n-1$ copies of the system. Importantly, it also holds that, for any Markovian process $\Vcal\in\Pbb^{M}_n$, we have $W(\Vcal(\Scal_1))=W(\Vcal(\Scal_2))$. Let $\Vcal$ be the closest Markovian process to $\Pcal$ according to the measure $N(\Pcal)$. Applying Lemma \ref{lem:cont} with $\tau=\Vcal(\Scal_1)=\Vcal(\Scal_2)$ then yields
\begin{align}
    |S(\Pcal(\Scal_1)||\gamma^{\otimes 2n-1})-S(\Pcal(\Scal_2)||\gamma^{\otimes 2n-1})&\le 2^{1/4}\qty(\ln 2+\frac{\ln \Tilde{m}^{-(2n-1)}}{\sqrt{2}})\qty(\sqrt{S(\Pcal(\Scal_1)||\Vcal(\Scal_1))}+\sqrt{S(\Pcal(\Scal_2)||\Vcal(\Scal_2))})^{1/2}\nonumber\\
    &\le 2^{1/4}\qty(\sqrt{2} \ln 2+(2n-1)S(\Pi_{E_{\textup{max}}}||\gamma))\qty[N(\Pcal)]^{1/4},
\end{align}
concluding the proof of Theorem 3.

\section{Maximum distillable work}

For a process $\Pcal\in\Pbb_n$ with system Hamiltonian $H_S$, let $\Scal\in\Sbb_n$ be the comb maximizing the work extractable from $\Pcal$, that is, $W^{\textup{comb}}(\Pcal)=W(\Pcal(\Scal))-W(\Scal)$. Note that $\Scal$ can be seen as a mapping from $n$ input states to $n+1$ output states. From the definition of $W(\Scal)$ we know the total work content $W_{\textup{out}}$ of its outputs is bounded by the work content $W_{\textup{in}}$ of its inputs plus its own work cost $W(\Scal)$. Then, $W(\Pcal(\Scal))$ being the work content of a single output (the last one), it must also be bounded by this quantity. Finally, since the inputs of $\Scal$ are $n$ systems with Hamiltonian $H_S$, $W_{\textup{in}}$ is upper bounded by $n$ times the maximum work content of a single system, namely, $F_{\textup{max}}\Def kTS(\Pi_{E_{\textup{max}}}||\gamma)$. This yields $W(\Pcal(\Scal))\le nF_{\textup{max}}+W(\Scal)$ or
\begin{equation}
    W^{\textup{comb}}(\Pcal)\le nF_{\textup{max}},
\end{equation}
thus providing an upper bound to the extractable work of any given process $\Pcal$.

Importantly, this bound may be saturated by a Markovian process. Let $\Qcal\in\Pbb_n^{M}$ be the Markovian $n$-step process consisting of $n$ independent channels having $\Pi_{E_{\textup{max}}}$ as fixed output. Then, by using sequential extraction and inputting only thermal states to the process, which can be done for free, one can extract exactly $nF_{\textup{max}}$ of work from this process, thus saturating the aforementioned bound.

\section{Tightness of the bounds}

Our bounds in Theorems~1--3 are all trivially saturated for Markovian processes. Finding examples of non-Markovian processes saturating the bounds of Theorems 1 and 3 might be particularly challenging, since they rely on the continuity bound of Lemma 1, which is generally not tight. However, here we show that the example given in Fig.~(2b) actually saturates the bound of Theorem 2. Since we already know that for this process $\Delta W^{\textup{MTC}}(\Pcal)=kT I(1:2)_{\rho_{12}}$, it suffices to show that $N(\Pcal)=I(1:2)_{\rho_{12}}$, where $\rho_{12}$ is the final global state from the example.

To calculate $N(\Pcal)= \min_{\Qcal\in\Pbb_n^{{M}}}\max_{\Scal\in\Sbb_n}S(\Pcal(\Scal)||\Qcal(\Scal))$ for the process $\Pcal$ of Fig. (2b), we first notice from Fig. (1a) that the last step of any control comb $\Scal$ is just a channel acting on system and ancilla, and from the contractivity of the relative entropy under channels we know the action of such channel can only decrease $S(\Pcal(\Scal)||\Qcal(\Scal))$, so it is enough to optimize over combs $\Scal^{\prime}$ without this last channel, like the one from Fig. \ref{fig:indexes}, consisting only of an initial system-ancilla state $\rho_{i_1i^{\prime}_1}$ and a subsequent channel $\Ecal_1$. Then, let $\Vcal\in\Pbb_n^{M}$ be the Markovian process consisting of two independent channels, the first one having $\rho_1$ as fixed output and the second one having $\rho_2$ as fixed output. Under the action of a comb $\Scal^{\prime}$ the output is $\Vcal(\Scal^{\prime})=\tilde{\Ecal}_1(\rho_1)\otimes\rho_2$, where $\tilde{\Ecal}_1(\sigma)=\tr_{i_2}\qty[\Ecal_1(\sigma\otimes\rho_{o^{\prime}_1})]$. Similarly, we have $\Pcal(\Scal^{\prime})=(\tilde{\Ecal}_1\otimes\Ical)\rho_{12}\Def \rho_{12}^{\prime}$, implying $S(\Pcal(\Scal^{\prime})||\Vcal(\Scal^{\prime}))=I(1:2)_{\rho_{12}^{\prime}}$, as both states have the same marginals, but only the second one has correlations. Note that $\max_{\Scal^{\prime}}S(\Pcal(\Scal^{\prime})||\Vcal(\Scal^{\prime}))=I(1:2)_{\rho_{12}}$ as the local processing given by $\tilde{\Ecal}_1$ cannot increase correlations in $\rho_{12}$, but they can always be preserved if $\tilde{\Ecal}_1$ is unitary, implying the distance between $\Pcal$ and $\Vcal$ is exactly $I(1:2)_{\rho_{12}}$. However, since $\Vcal$ is not necessarily the closest Markovian process to $\Pcal$, we can only state $N(\Pcal)\le I(1:2)_{\rho_{12}}$. Finally, we use Eq. \eqref{eq:N-I}, which in this context implies $N(\Pcal)\ge I(1:2)_{\rho_{12}}$, to conclude that $N(\Pcal)= I(1:2)_{\rho_{12}}$ for the process of Fig. (2b), thus saturating the bound of Theorem 2.

\end{document}